\documentclass[11pt]{article}
\usepackage{amssymb}
\usepackage{amsfonts}
\usepackage{latexsym}

\usepackage{amsmath}
\usepackage{graphicx}

\def\beq{\begin{eqnarray}}
\def\eeq{\end{eqnarray}}
\def\ln{\,\mbox{ln}\,}

\def\al{\alpha}
\def\be{\beta}

\def\ga{\gamma}

\def\pa{\partial}

\def\si{\sigma}

\def\ph{\varphi}

\def\th{\theta}

\def\Ga{\Gamma}

\def\La{\Lambda}

\begin{document}

\begin{center}

{\large\sc Effective field theory and fundamental interactions}
\vskip 6mm

{\bf Ilya L. Shapiro}
 \footnote{Electronic address: shapiro@fisica.ufjf.br

On leave from Tomsk State Pedagogical University, Tomsk, Russia}
\vskip 2mm

{\small\sl Departamento de F\'{\i}sica -- ICE, 
Universidade Federal de Juiz de Fora, MG, Brazil}
\end{center}

\vskip 6mm

\begin{quotation}

\noindent
{\large \sl Abstract.}
There are many possible gravitational applications of an 
effective approach to Quantum Field Theory (QFT) in curved 
space. 
We present a brief review of effective approach and discuss 
its impact for such relevant issues as the cosmological 
constant (CC) problem and inflation driven by vacuum quantum 
effects. Furthermore it is shown how one can impose 
significant theoretical constraints on a non-metric gravity 
using only theoretical effective field theory framework.
\end{quotation}

\section{Introduction}

$\,$ \quad 
The effective approach to QFT implies the 
low-energy phenomena being described independently on the
(sometimes unknown) fundamental physics. A famous example 
is the low-energy QCD, where the Chiral Perturbation Theory 
helps to achieve results fitting both lattice simulations 
and experiment, in a situation when usual perturbative methods 
are not applicable. There are many good reviews on the 
effective approach, see e.g. \cite{manohar} and references
therein. 

Sometimes, the effective approach enables one to 
reduce the requirements to a theory, e.g. extract relevant 
low-energy information even from the non-renormalizable 
theories, where the high-energy UV limit is problematic.
It might happen that this kind of considerations is applicable 
even to quantum gravity, e.g. one can extract from this theory 
the long-distance quantum correction to the Newton potential 
\cite{don}. The theoretical relevance of this result can not 
be questioned (even in spite of the extreme smallness 
of the effect), because it helps in better understanding of 
the consistency conditions for the effective approach,
e.g. the importance of the full set of Feynman diagrams 
\cite{anhesh,don2}. However, despite the theoretical 
investigation of Quantum Gravity is really fascinating, we 
can not be sure that they have something to do with Nature. 
It might happen that 
the gravitational field should not be quantized at all, 
representing just a classical background for the quantized
matter fields and particles.

In the present review we shall concentrate on the effective
approach to QFT in curved space. This is 
perhaps the most natural way of investigating quantum and
gravitational phenomena together \cite{birdav,book}. The enormous 
success of QFT in describing the theories like QED 
or Standard Model prove the validity of QFT methods. 
On the other hand, since there are many indications that 
our space-time is curved, the QFT in a curved space is, let 
us say, a correct theory applied in a correct place. Our 
main point is that, using the 
effective approach and extracting relevant information 
at different energy scales, one has a chance to learn a natural 
description for many interesting phenomena, such as inflation
or scale (or time) dependence of the vacuum energy.
Moreover, while making quantitative predictions is 
sometimes difficult yet, one can exclude some  definite 
options for the space-time using effective approach. 

The important aspect of effective approach is the notion of 
decoupling. At classical level decoupling means that a
heavy field doesn't propagate at low energies. This can be 
easily seen observing the propagator of a massive 
particle at low energy
\beq
\frac{1}{k^2+M^2}\,\approx\,
\frac{1}{M^2}
\,+\,{\cal O}\left( \frac{k^2}{M^4} \right)\,,
\qquad k^2 \ll M^2\,.
\label{propagator}
\eeq
For example, if we take a gravity theory with higher derivatives,
the typical value of $M$ would be $M_P$. Already for $|k|=1\,TeV$
the ratio $k^2/M^2 \propto 10^{-32}$. It is clear that only 
massless mode of gravity is relevant even for the early Universe
or other potentially observable phenomena. The important 
exception is inflation, which will be discussed below.

The decoupling theorem explains how 
the suppression of the effects of heavy particles occurs at 
quantum level. Consider the QED example in flat space.
The 1-loop vacuum polarization is 
\beq
-\frac{e^2\,\th_{\mu\nu}}{2\pi^2}\,\int_0^1 dx \, x(1-x)
\,
\ln \frac{m_e^2+p^2x(1-x)}{4\pi \mu^2},
\eeq
where
$\,\th_{\mu\nu}=(p_{\mu}p_{\nu}-p^2g_{\mu\nu})$ and
$\mu$ is the parameter of dimensional regularization. The minimal 
subtraction scheme of renormalization results in the $\mu$-dependence
for the relevant observables. For instance, the 
$\be$-function $\beta^{\overline{\rm MS}}$ results after acting 
$\,\,\frac{e}{2}\mu \frac{d}{d\mu}\,\,$ on the formfactor of 
$\,\th_{\mu\nu}$
\beq
\beta_e^{\overline{\rm MS}}\,\,=\,\,\frac{e^3}{12\pi^2}.
\eeq
This $\be$-function can not tell us about the decoupling and 
possesses an artificial universality, because the 
${\overline {\rm MS}}$ scheme is efficient only in the UV limit.
In order to define physical $\be$-function, let us apply the 
mass-dependent renormalization scheme. Subtracting the divergence 
at $\,p^2=M^2\,$ and taking derivative 
$\,\frac{e}{2}M \frac{d}{dM}\,$ we arrive at 
$$
\beta_e\,=\,\frac{e^3}{2\pi^2}\int_0^1
\, dx\, x(1-x)\,\frac{M^2x(1-x)}{m_e^2+M^2x(1-x)}.
$$
The UV limit ($M \gg m_e$) gives 
$\quad \beta_e = \beta_e^{\overline{\rm MS}}$ and in the 
the IR limit ($M \ll m_e$) we meet a quadratic decoupling
(Appelquist \& Carazzone theorem \cite{AC})
\beq
\beta_e \, = \, \frac{e^3}{60\pi^2}\,\cdot\,
\frac{M^2}{m_e^2}
\,\,+\,\,{\cal O}\left(\frac{M^4}{m^4}\right)\,.
\label{QED}
\eeq
Compared to $\,\be_e^{UV}=\be_e^{\overline{\rm MS}}$,
in the IR there is a suppression of the quantum contribution.

\section{Decoupling and cosmological constant}

$\,$\quad
Quantitative investigation of decoupling for quantum 
matter on curved background started recently \cite{apco}. 
Why this is  interesting and important? The most explicit 
example is the decoupling of the quantum contributions to
the CC (vacuum energy) \cite{CCfit}. 
Let us assume that the Appelquist \& Carazzone-like quadratic 
decoupling holds for a CC and consider 
the consequences for the present-day Universe.
One has to associate the scale \ $\mu \equiv H$
with the Hubble parameter \cite{nova}. Let us notice that 
this identification of the scale has serious advantages 
over other possible choices \cite{CC,reuter}.

Remember that, in the ${\overline{\rm MS}}$ scheme, 
$\be_\La \sim m^4$, $m$  being mass of a quantum matter 
field \cite{book}. Then the quadratically suppressed 
expression is \cite{CCfit} 
\beq
H\,\frac{d\Lambda}{dH}\,=\,
\beta_{\Lambda}\,=\,
\sum\limits_i\,c_i\,\frac{H^2}{m_i^2}\,\times m_i^4
\,=\,\frac{\si}{(4\pi)^2}\,M^2\,H^2\,,
\label{IR CC}
\eeq
where $M\,$ is an unknown mass parameter and $\,\si=\pm 1\,$.
The expression $\,\si M^2\,$ is the algebraic
sum of the contributions of all virtual fields, the ones 
of the heaviest particles being the most relevant. The sign
$\,\si\,$ depends on whether fermions or bosons dominate
in the particle spectrum.

Let us assume, for a moment, that $\,M^2=M_P^2$. Then 
we find 
\beq
|\beta_{\Lambda}|\sim 10^{-47}\,GeV^4\,,
\label{47}
\eeq
that is close to the existing supernovae and CMB data
for the vacuum energy density. Therefore, the renormalization 
group may, in principle, explain the variation of the 
vacuum energy with the change of the scale (which, in 
presence of matter, means also variation of time) without 
introducing special entities like quintessence. 

It is clear that the renormalization group can not solve 
the famous CC problem \cite{weinberg89},
neither the coincidence problem. The fine-tuning of the
vacuum CC is performed at the instant 
when we define the initial point $\La_0$ of the 
renormalization group 
flow for $\La$ \cite{nova}. However, the running 
of the CC may perform in the range comparable to the 
observed CC, making the coincidence problem less severe.

The example of a cosmological model with running CC has 
been developed
in \cite{CCfit}. Along with the renormalization group 
equation (\ref{IR CC}), there is also Friedmann equation
(for simplicity we consider $k=0$ here)
\beq
H^{2}=\frac{8\pi G }{3}\left( \rho +\Lambda\right)\,,
\label{Friedmann}
\eeq
and the conservation law. The last can be chosen in many 
possible ways. The choice 
\beq
\frac{d\La}{dt}\,+\,\frac{d\rho}{dt}
\,+\,3H\,(\rho+P)\,=\,0
\label{conservation}
\eeq
means we admit the energy exchange between the vacuum and 
matter sectors (see also \cite{waga}, where similar 
cosmological model with $\La_0=0$ have been developed 
earlier). The solution of the equations (\ref{IR CC}),
(\ref{Friedmann}), (\ref{conservation}) is analytical
for any equation of state $P=\al \rho$.
In terms of the red-shift parameter $\,z=a_0/a-1$ this 
solution has the form
\beq
\rho(z;\nu) \,=\,\rho_0\,(1+z)^{3(1-\nu)(\al+1)}
\,,\qquad
\Lambda(z;\nu)\,=\,\Lambda_0
\,+\,\frac{\nu}{1-\nu}\,\left[\,\rho(z;\nu)-\rho_0\,\right]\,,
\label{1}
\eeq
where
$\,\,\rho_0,\, \Lambda_0\,\,$ are the present day values,
and \ $\nu\,=\,\sigma\,M^2/12\pi M_P^2\,.$
When $\nu\rightarrow 0$
we recover the standard result for $\,\La=const$.

The value of $|\nu|$ has to satisfy the constraint 
$|\nu|\ll 1$, for otherwise there is a problem with the 
nucleosynthesis \cite{CCfit}.
The ``canonical'' value $\,M^2=M_P^2\,$ gives
\ $|\nu|=\nu_0\,\simeq\, 2.6\times 10^{-2}$,
\ compatible with this constraint.
The next question is whether the permitted values $\nu \ll 1$ 
may lead to observable consequences. The answer is positive.
For example, the relative deviation 
$$
\delta_{\Lambda}(z;\nu)=\frac{
{\Lambda}(z;\nu)-{\Lambda}_0}{{\Lambda}_0}
$$
can be evaluated in the linear in $\nu$ approximation
with respect to the red shift $z_0$ corresponding 
to the existing supernovae data \cite{SN}
\beq
\delta_{\Lambda}(z;\nu)=\,\frac{\nu
\,\Omega_M^0}{\Omega_{\Lambda}^0}
\left[(1+z)^3-(1+z_0)^3\right].
\label{3}
\eeq
Taking $z_0\simeq 0.5$ with $\,\Omega_M^0=0.3\,$
and $\Omega_{\Lambda}^0=0.7$,
and $\nu=\nu_0$, we find
\ $\delta_{\Lambda}(z=1.5;\nu_0)\ \approx \ 14\%$, \
that is a potentially measurable effect. It is 
remarkable that the cubic relation (\ref{3}) holds also for 
another choice of the conservation law \cite{gruni}.

In fact, the QFT methods can teach us more lessons about the 
form and origin of quantum corrections to the CC 
\cite{Note}. Let us remind that behind the renormalization 
group there is a well-defined object called Effective Action
of vacuum (EA)
\beq
e^{i\Ga[g_{\mu\nu}]}\,=\,
\int {\cal D}\Phi\,e^{iS[\Phi\,,\,g_{\mu\nu}]}\,.
\label{4}
\eeq
Typically, $\Ga[g_{\mu\nu}]$ is a complicated non-local 
functional of the background metric. The renormalization 
group reflects the scaling dependence of EA. When 
considering the low-energy cosmological applications, we 
safely can 
perform the expansion in the Hubble parameter $H$ similar 
to the one of (\ref{propagator}). Let us remark that the 
masses of all particles are many orders of magnitude smaller 
then the present day value $H_0\approx 10^{-42}\,GeV$. E.g.
the neutrino masses are about $10^{-12}\,GeV$ while the 
QCD scale $\La_{QCD}\approx 10^{-2}\,GeV$. Hence the $H$ 
expansion is just a reliable way to parametrize all 
non-localities including the ones related to the non-perturbative
QFT effects (non-perturbative in couplings but not in $H$)
\cite{CCfit}.

At present, our ability of deriving an explicit form of EA
for the massive fields is very restricted. But we are certain
it is a covariant functional \cite{book}, and even this 
knowledge is sufficient for obtaining an essential piece 
of information. Due to covariance,
$\Ga[g_{\mu\nu}]$ must be even in metric derivatives. 
In the cosmological setting this means that the 
linear in  $\,H\,$ quantum corrections to the CC are 
completely ruled out. Of course, the last 
statement is valid only for the case when metric is the 
unique
background field. If there is an extra light (with the mass
comparable to $H_0$) field \ $\ph$, such as quintessence, then 
the EA is $\Ga[g_{\mu\nu},\ph]$ and any kind of a
functional dependence between vacuum energy and $H$ may 
take place. However, without quintessence no QFT can 
produce \ ${\cal O}(H)$  \ quantum corrections. Let us 
stress that this restriction is valid independently on the 
nature of quantum corrections (local, non-local, perturbative 
or non-perturbative ones).

Without quintessence the quantum corrections start from 
$H^2$ and therefore, e.g. the QCD vacuum effects play no role
in cosmology. The relevant quantum effects on the 
vacuum energy may come only from the Planck scale physics or, 
at least, from the GUT-scale physics.  

\section{Higher derivative sector}

In order to derive decoupling theorem for gravity we have 
considered massive fields on the classical metric background
\cite{apco}. Unfortunately there is no completely covariant 
technique compatible with the mass-dependent renormalization 
schemes. Hence we can perform calculations only for the
linearized gravity on the flat background
$\,g_{\mu\nu}=\eta_{\mu\nu} + h_{\mu\nu}$.
The corrections to the graviton propagator come from the 
Feynman diagrams at the Figure 1.

\vskip 2mm
\begin{tabular}{c}
\mbox{\hspace{+1.0cm}}
\includegraphics[width=12cm,angle=0]{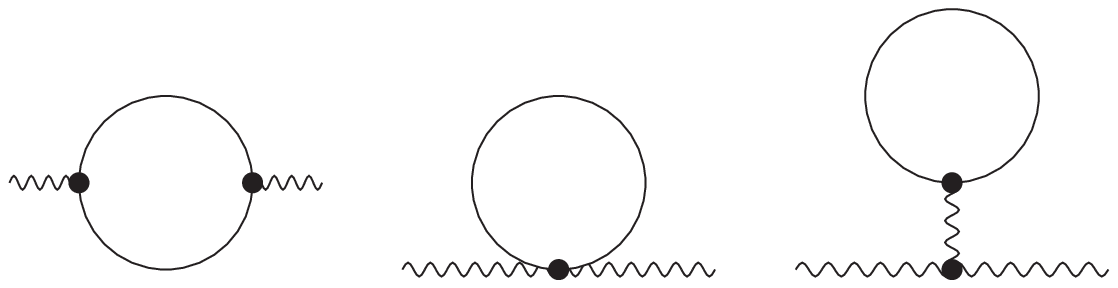}
\end{tabular}
\begin{quotation}
{\bf Figure 1.} 
{\sl Wavy lines mean $h_{\mu\nu}$ and solid lines quantum 
matter field.}
\end{quotation}

The polarization operator must be compared to the
tensor structure of the Lagrangian
\beq
L = - \La
-\frac{1}{16\pi G}\,R + a_1C^2 + a_2E + a_3{\Box}R + a_4 R^2\,,
\label{L}
\eeq
where $C^2=C^2_{\mu\nu\al\be}$ is the square of the Weyl 
tensor and $E$ is the integrand of the Gauss-Bonnet 
topological invariant $\,E=R^2_{\mu\nu\al\be}-4R^2_{\mu\nu}+R^2$.

An alternative equivalent way of calculation is using the heat 
kernel solution in the second order in curvature approximation
\cite{avramidi,bavi90}.

For the formfactors we find, e.g. for the real scalar field
\beq
k_\La \,=\, \frac{3m^4}{8\,(4\pi)^2}\,,
\qquad
k_R \,=\,\frac{m^2}{2\,(4\pi)^2}\,
\left( \xi-\frac16 \right)\,,
\qquad
k_1(a)\, = \,\frac{8A}{15\,a^4}
\,+\,\frac{2}{45\,a^2}\,+\,\frac{1}{150}\,,
\label{form}
\eeq
where 
$$
A = 1 + \frac{1}{a}\ln
\left|\frac{2-a}{2+a}\right|
\,,\quad
a^2=\frac{4\Box}{4m^2 -\Box }.
$$
Obviously, constant formfactors mean zero $\be$-functions 
for the CC and $G$. At the same time for the coefficient of the 
Weyl term we find
\beq
\be_1\,=\, -\,\frac{1}{(4\pi)^2}\, \left(\,
\frac{1}{18a^2}\,-\,\frac{1}{180}\,-\,
\frac{a^2-4}{6a^4}\,A\,\right)\,.
\label{beta 1}
\eeq
Then
\beq
\be_1^{UV}=-\frac{1}{(4\pi)^2}\frac{1}{120}
+ {\cal O}\left(\frac{m^2}{p^2}\right)\,,
\qquad
\mbox{and}
\qquad
\be_1^{IR}\,=\,-\,\frac{1}{1680\,(4\pi)^2}\,\cdot\,
\frac{p^2}{m^2}
\,+\,{\cal O}\left(\frac{p^4}{m^4}\right)\,.
\label{beta 1 IR}
\eeq
The last formula is nothing but the 
Appelquist \& Carazzone theorem for  gravity.
Our calculations \cite{apco} have shown it holds
in all higher derivative sector, including the theories
with the Spontaneous Symmetry Breaking \cite{sponta}.

An expansion \ $g_{\mu\nu}=\eta_{\mu\nu}+h_{\mu\nu}$
works well for higher derivative terms, but not
for $\La$ and  $G$. Why did we obtain
$\beta_{\Lambda}=\beta_{1/G}\equiv 0$? In fact, running 
means the presence of a $f(\Box)=\ln(\Box/\mu^2)$-like
formfactor. In QED, in the UV limit we meet the term
$$
-\frac{e^2}{4}F_{\mu\nu}F^{\mu\nu}
+ \frac{e^4}{3(4\pi )^2}\,F_{\mu\nu}\,
\ln \left(-\frac{\Box}{\mu^2}\right)\,F^{\mu\nu}\,.
$$
Similarly in gravity it is possible to insert
$$
C_{\mu\nu\al\be}\,f(\Box)\,C^{\mu\nu\al\be}
\quad
\mbox{or}
\quad\quad
R\,f(\Box)\,R
$$
in the higher derivative sector. However, no insertion is 
possible for $\La$ and $1/G$, since $\Box\Lambda=0$ and
$\Box R\,$ is a total derivative.

Does it mean that $\,\be_\La$ and $\be_{1/G}\,$ really 
equal zero? From my point of view the answer is negative, 
for otherwise we meet a
divergence between the mass-dependent renormalization 
scheme and $\,{\overline{MS}}$-scheme in the UV where 
they are supposed to be the same. 
Perhaps calculations on a flat background are not 
appropriate for deriving the renormalization group  
equations for $\La$ and $1/G$. This hypothesis is quite 
natural, especially because flat space is not a classical 
solution in the presence of the CC.

\section{Anomaly-induced inflation}

At present, we are not able to derive or disprove the 
relation (\ref{IR CC}) and hence the cosmological model 
with variable CC described in section 2 is 
essentially phenomenological. What we can calculate is
the decoupling in the higher derivative sector. In fact,
this is also very interesting, 
creating a solid basis for the anomaly-induced inflation
(modified Starobinsky model) \cite{asta,shocom}.

The most important result is the $\be_3$-function in the 
theory with broken supersymmetry. Due to the decoupling of
the heavy sparticles this $\be$-function smoothly changes 
its sign from negative in the UV to positive in IR \cite{apco}. 
The $\be_3$-function is nothing else but the coefficient $c$
in the expression for the conformal anomaly 
\beq
T = <T_\mu^\mu> \,=\, - \,(wC^2 + bE + c\,{\Box} R)\,,
\label{anomaly}
\eeq
where $w,b,c$ depend on the number of fields of different 
spins $\,N_0,\,N_{1/2},\,N_1$ in the underlying GUT. The 
signs of other two coefficients $w>0$ and $b<0$ are universal 
such that only the sign of $c$ alternates. 

Taking (\ref{anomaly}) into account we arrive at the equation 
for the conformal factor of the metric 
\beq
\frac{{\ddot {\ddot a} }}{a}
+\frac{3{\dot a} {\dot {\ddot a}} }{a^2}
+\frac{{\ddot a}^2}{a^2}
-\left( 5+\frac{4b}{c}\right)
\frac{{\ddot a}{\dot a}^2}{a^3}
-\frac{M_{P}^{2}}{8\pi c}
\left( \frac{{\ddot a}}{a}+
\frac{{\ddot a}^2}{a^{2}}
-\frac{2\Lambda }{3}\right)\,=\,0\, ,
\label{equation}
\eeq
The important particular solutions are
\beq
a(t) \,=\, a_0 \,e^{Ht}\,,
\qquad
H^2\,=\, - \frac{M^2_P}{32\pi b} \, \Big(\,1\pm \,
\Big[\,1+\frac{64\pi b\,\Lambda }{3\,M_P^2}
\,\Big]^{1/2}\,\Big)\,.
\label{H}
\eeq
For \ $0<\La \ll M_P^2$ \ there are two solutions:
\beq
H_f\,\approx\,\sqrt{{\Lambda }/{3}}\,,
\qquad \mbox{and}\qquad
H_i\,\approx\, \sqrt{- \frac{M^2_P}{16\pi b}
- \frac{\Lambda }{3}}
\approx \frac{M_P}{\sqrt{- 16\pi b}}\,,
\label{IR-UV}
\eeq
the last one corresponds to the Starobinsky inflation
\cite{star}. The inflationary solution is stable under
perturbations of $\,\si(t)=\ln a(t)\,$ in case the 
condition 
\beq
c \,\,\sim\,\,
\frac{N_0}{18} + \frac{N_{1/2}}{3} - N_1 > 0
\label{c}
\eeq
is satisfied \cite{star,asta}. If the gauge model includes 
many fermions and scalars for a given number of vectors, its
vacuum quantum effects lead to the stable inflation, otherwise
the inflation is unstable and there is a chance of a graceful 
exit to the FRW-like evolution. The original Starobinsky model
\cite{star} is based on the unstable case and involves heavy 
fine-tunings for the initial data. Our purpose is to avoid 
fine-tunings at all, fortunately the effective approch shows 
this is possible.

In order to understand how one can consider both stable and
unstable inflations in the very same model, let us assume
that at the UV $\,(H \gg M_F)\,$ there is supersymmetry, 
e.g. MSSM with a particle content
$$
N_1=12\,,\,\,\,\,\,\,\,\,\,\,\,\,\,\,\,\,
N_{1/2}=32\,,\,\,\,\,\,\,\,\,\,\,\,\,\,\,\,\,
N_0=104\,.
$$
This provides stable inflation, because $c>0$. Similar 
situation holds for any realistic SUSY model. The advantage 
of stable inflation is that it is independent on the initial 
data. But why should inflation end?

Already at the MSM $\,(\,N_{1,\,1/2,\,0}=12,24,4)$ scale 
$\,H \sim 10^2\,GeV\,$ the 
inflation is unstable, \ $c<0$. One can suggest the 
following physical interpretation of this sign difference.
We know that all sparticles are heavy, hence they
decouple, when $\,\,H\,\,$ becomes smaller than their 
masses. According to \cite{apco},
the transition from \ $c>0$ \ to \ $c<0$ \ is smooth,
giving a hope for a graceful exit. 

The next question is why the magnitude of the Hubble parameter 
is becoming
smaller in the course of inflation. An explicit calculation 
of the contribution of a particle of mass $m$ in the regime 
$H \gg m$ has been performed in \cite{shocom}. Extrapolating 
the result until $H \approx m$ we arrive at the curve
$\si(t)$ shown at the Figure 2.  
\vskip 2mm

\begin{tabular}{c}
\mbox{\hspace{+3.0cm}}
\includegraphics[width=8cm,angle=0]{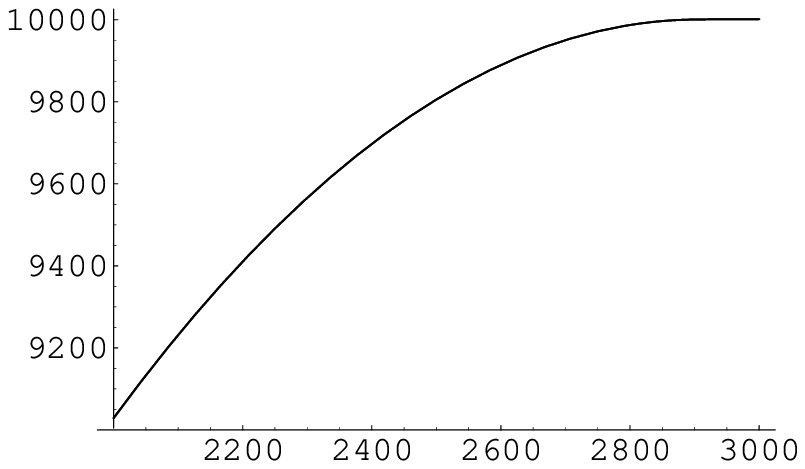}
\end{tabular}

\begin{quotation}
{\bf Figure 2.} \ {\sl The plot of $\si(t)=\ln a(t)$. 
An approximate analytic expression is 
$\,\si = H_i\,t\,-\,H^2_i\tilde{f}t^2/4\,$.
The value of $\tilde{f}$ depends on the mass spectrum 
of the theory. For MSSM the total amount of $e$-folds
is as large as $10^{32}$,  but only $65$
last ones, where $\,H\propto M_*$ (SUSY decoupling 
scale) are relevant.}
\end{quotation}
\vskip 1mm

An additional advantage of the anomaly-induced inflation
becomes clear when investigating 
the gravitational waves in this model. It turns out that, 
during the last 65 $e$-folds of inflation the production 
of these waves is restricted and they almost do not amplify
\cite{wave,asta} (see also \cite{Hawking}). 
Again, this nice feature follows without 
any artificial restrictions or fine-tuning.

As we can see, the anomaly-induced model represents a 
strong candidate to the role of natural inflation scenario, 
because it does not require fine-tuning for initial data 
and does not need a special entity such as inflaton. 
However, small definite information is available
about the most important intermediate stage of inflation. 
In order to 
obtain this information one needs further development 
of QFT in curved space-time, mainly investigating the 
vacuum contributions of massive fields on a dynamical 
metric background.

\section{Restrictions on Space-Time Geometry
from effective QFT}

Consider how one can impose the restrictions on a Space-Time 
Geometry using effective approach to QFT. In fact, we can 
impose very rigid constraints on the propagation of a 
space-time torsion in the effective framework 
\cite{betor,guhesh,torsi}. 

The theories of gravity with torsion attracted significant 
attention for a long time (see e.g. \cite{hehl,torsi} and
references therein). Torsion $\,T^\alpha_{\;\beta\gamma}\,$
is independent (on metric) characteristic of a space-time 
manifold which is defined by the relation
$$
{\Gamma}^\alpha_{\;\beta\gamma} -
{\Gamma}^\alpha_{\;\gamma\beta} =
T^\alpha_{\;\beta\gamma}\,.
$$
It is useful dividing torsion into irreducible components
\beq
T_{\alpha\beta\mu} =
\frac{1}{3} \left( T_{\beta}g_{\alpha\mu} -
T_{\mu}g_{\alpha\beta} \right)
- \frac{1}{6} \varepsilon_{\alpha\beta\mu\nu}
\,S^{\nu} + q_{\alpha\beta\mu}\,.
\label{t1}
\eeq
Interaction to the Dirac fermion has the form (we assume, for 
simplicity, the flat metric)
\beq
{\cal L} = i\bar{\psi}\gamma^\mu(\partial_\mu
+i\eta_1\gamma^5S_\mu+i\eta_2T_\mu)\psi+m\bar{\psi}\psi\,,
\label{t2}
\eeq
where $\eta_1,\,\eta_2$ \ are nonminimal parameters.
The minimal case corresponds to $\eta_1 = 1/8,\;\;\eta_2 = 0$,
so the presence of torsion means, at least, that any fermion 
is coupled to an axial vector $S_\mu$
\beq
S = i\int d^4 x\,\bar{\psi}\,\ga^\mu
\,\left(\partial_\mu + i\eta_1\ga^5\,S_\mu-im
\right)\,\psi\,.
\label{t3}
\eeq
Let us notice that the interaction with torsion is 
nothing else but the known CPT \& Lorentz violating term
\cite{Kost}.
The question is whether we can construct a QFT
for \ $\psi$ and $S_\mu$ which would be
unitary and also renormalizable as an effective QFT.
Let us consider this QFT in two steps. 
First, quantizing $\psi$ we meet the two types of divergences
\cite{book,betor}:
\beq
S_{\mu\nu}\,S^{\mu\nu}
\qquad \mbox{and} \qquad
m^2 S_\mu\,S^\mu\,,\qquad \mbox{where} \qquad
S_{\mu\nu}=\pa_\mu S_\nu - \pa_\nu S_\mu\,.
\label{t4}
\eeq
It is well-known that unitarity forbids simultaneous 
$S_\mu^\bot$ and $S_\mu^{\|}$ propagation. Therefore,
the unique possibility for a dynamical torsion is 
\cite{betor}
\beq
S_{tor} = \int d^4x\,\left\{ - \frac14\,S_{\mu\nu}^2
+ M^2 S_\mu^2\right\}\,.
\label{t5}
\eeq

As a second step we have to investigate whether 
the effective quantum theory of fermion coupled to
dynamical torsion is consistent. An extremely involved 
calculations yield \cite{guhesh} a longitudinal \
$\left(\pa_\mu S^\mu\right)^2$-type divergence
at the two-loop level.
This means there is a severe conflict between 
unitarity and renormalizability in the low-energy 
corner of the theory.

One possible solution is to assume \ $\eta^4\,m^2 \ll M^4$.
This means that either $M \gg m$ for all fermions, or that 
\ $\eta$ \ is extremely small. In both cases there is no 
chance to observe propagating torsion. The lower bound 
from LEP is $M/\eta \leq 3\,TeV$ \cite{betor}, 
that is far beyond the necessary condition presented 
above. Finally, torsion can be a composite field (e.g. it can 
be a vacuum condensate) but it can not be an independent 
and experimentally observable propagating field. 

\section{Conclusions}

\qquad 
The effective approach to QFT
in curved space-time may tell us a lot about
gravitational physics, especially in cosmology.
Perhaps the most interesting problem is evaluation
of vacuum effective action for massive quantum
fields. Working in this direction one may prove or disprove
the possibility of a time-dependent cosmological constant.
The same calculation is vital for further theoretical 
development of the anomaly-induced inflation model. 

Within the existing formalism of QFT in curved space
and known calculational techniques we can learn something 
relevant about the possible form of quantum corrections, 
e.g. exclude the ${\cal O}(H)$-type corrections to the CC. 
If this 
kind of dependence will be someday detected, it will be 
direct indication to the existence of a qualitatively 
new field such as quintessence. 

Surprisingly, one can exclude some relevant options (such 
as propagating torsion) for the space-time geometry using 
QFT methods and effective approach.

\vskip 12mm
\noindent
{\Large\bf Acknowledgements.} 
This short review has been based on 
the works done in collaboration with M. Asorey, J. Fabris,
E. Gorbar, J. A. Helay$\ddot{\rm e}$l-Neto, J. Sol\`a, 
G. de Berredo Peixoto and Ana Pelinson. I am very grateful 
to all of them. The work has been partially supported by 
CNPq, FAPEMIG and ICTP. 


\end{document}